\documentclass[aps,prb,twocolumn,superscriptaddress,longbibliography]{revtex4-1}
\usepackage{graphicx}
\usepackage{latexsym}
\usepackage{amssymb}
\usepackage{amsmath}
\usepackage{amsfonts}
\usepackage{upgreek}
\usepackage{float}
\usepackage{bm}
\usepackage{multirow}
\usepackage{color}
\usepackage[T1]{fontenc}
\usepackage{hyperref}
\hypersetup{
colorlinks = true,
linkcolor = [rgb]{0.70,0.13,0.13},
citecolor = [rgb]{0.13,0.55,0.13},
urlcolor  = [rgb]{0.25, 0.41, 0.88}}
\newcommand{\ket}[1]{|#1\rangle}
\newcommand{\bra}[1]{\langle#1|}
\newcommand{\braket}[2]{\langle#1|      #2\rangle}

\begin{document}

\title{Biorthogonal quantum criticality in non-Hermitian many-body systems}
\author{Gaoyong Sun}
\thanks{Corresponding author: gysun@nuaa.edu.cn}
\affiliation{College of Science, Nanjing University of Aeronautics and Astronautics, Nanjing, 211106, China}
\affiliation{Key Laboratory of Aerospace Information Materials and Physics (Nanjing University of Aeronautics and Astronautics), MIIT, Nanjing 211106, China}

\author{Jia-Chen Tang}
\affiliation{College of Science, Nanjing University of Aeronautics and Astronautics, Nanjing, 211106, China}
\affiliation{Key Laboratory of Aerospace Information Materials and Physics (Nanjing University of Aeronautics and Astronautics), MIIT, Nanjing 211106, China}

\author{Su-Peng Kou}
\thanks{Corresponding author: spkou@bnu.edu.cn}
\affiliation{Center for Advanced Quantum Studies, Department of Physics, Beijing Normal University, Beijing 100875, China}

\begin{abstract}
We develop the perturbation theory of the fidelity susceptibility in biorthogonal bases for arbitrary interacting non-Hermitian many-body systems with real eigenvalues. 
The quantum criticality in the non-Hermitian transverse field Ising chain is investigated by the second derivative of ground-state energy and the ground-state fidelity susceptibility.
We show that the system undergoes a second-order phase transition with the Ising universal class by numerically computing the critical points and the critical exponents from the finite-size scaling theory. 
Interestingly, our results indicate that the biorthogonal quantum phase transitions are described by the biorthogonal fidelity susceptibility instead of the conventional fidelity susceptibility. 

\end{abstract}

\maketitle

\section{Introduction}
The study of quantum matters and quantum phase transitions is one of the central parts in condensed matter physics \cite{sachdev1999quantum}. 
For conventional Hermitian many-body systems, a quantum phase transition is usually characterized by a qualitative change in the ground-state eigenfunction 
and the non-analyticity of the ground-state energy at the critical point in thermodynamic limit \cite{sachdev1999quantum}.
The corresponding quantum state of matter can be distinguished by the order parameters or the topological quantities \cite{levin2006detecting}.
Moreover, the nature of phase transitions (or the critical exponents) can be described and obtained by the finite-size scaling theory \cite{Fisher1972,Fisher1974}.

Non-Hermitian systems that can be realized by a gain and loss process or by a nonreciprocal hopping exhibit many intriguing unique phenomena 
beyond Hermitian systems \cite{bergholtz2021exceptional, ashida2020non},
for example, the breakdown of the bulk-boundary correspondence and the non-Hermitian skin effect \cite{lee2016anomalous,yao2018edge,kunst2018biorthogonal,xiong2018does,
gong2018topological,alvarez2018non,yokomizo2019non,okuma2020topological,zhang2020correspondence,yang2020non,
wang2020defective,jiang2020topological,weidemann2020topological,xiao2020non,borgnia2020non}, 
exceptional points and bulk Fermi arcs \cite{heiss2012physics,kozii2017non,hodaei2017enhanced,zhou2018observation,miri2019exceptional,park2019observation,yang2019non,
ozdemir2019parity,dora2019kibble,zhang2019high,jin2020hybrid,Xiao2020Observation}, 
phase transitions without gap closing \cite{matsumoto2020continuous,yang2020anomalous}, etc.
New theories or concepts, i. e. non-Bloch band theory \cite{yao2018edge,yokomizo2019non,zhang2020correspondence}, usually are in demand to understand such non-Hermitian phenomena.
Recently, non-Hermitian many-body physics were explored to consider the interplay of the interaction and the non-Hermiticity \cite{jin2013scaling,matsumoto2020continuous,yang2020anomalous,
ashida2017parity,herviou2019entanglement,chang2019entanglement,mu2020emergent,lee2020many,pan2020non,pan2020interaction,xu2020topological,zhang2020skin,
lee2020many2,shackleton2020protection,liu2020non,yang2021exceptional,hanai2019non,hamazaki2019non,xi2021classification,yamamoto2019theory,hanai2020critical}.
One central issue is to understand the phase transition and the quantum criticality \cite{yao2018edge,ashida2017parity,dora2019kibble,hanai2019non,hamazaki2019non,xi2021classification,yamamoto2019theory,hanai2020critical,arouca2020unconventional}.
However, the study of non-Hermitian many-body systems is extremely difficult because of the complexity of many-body systems 
and the demand of the high numerical accuracy (i. e. the quadruple precision is required even for single-particle computations \cite{yao2018edge}).

Fidelity (or fidelity susceptibility (FS)), a simple concept from quantum information, is widely used to detect quantum phase transitions in Hermitian many-body systems 
\cite{Zanardi2006, Venuti2007, You2007,Albuquerque2010, Gu2010,Sun2017,Zhu2018,Wei2018,Wei2019,Chen2008,Gu2008,Yang2008,
Kwok2008,Gong2008,Yu2009,Schwandt2009,Lu2018,Rams2011,Li2012,Victor2012,Damski2013,Carrasquilla2013,Lacki2014,Sun2016,Yang2007,
Fjestad2018,Langari2012,Sun2015,cincio2019universal,sun2019fidelity}.
Recently, fidelity susceptibility has been generalized to the non-Hermitian systems to characterize non-Hermitian phase transitions \cite{jiang2018topological,matsumoto2020continuous,yang2020anomalous,
wang2020effective,guo2020non,nishiyama2020imaginary,nishiyama2020fidelity,tzeng2021hunting, solnyshkov2021quantum}.
Because there exist two sets of eigenstates (left and right eigenstates) \cite{brody2013biorthogonal}, 
one can define two types of fidelities depending on the usage of left and right eigenstates \cite{herviou2019entanglement}.
For non-Hermitian systems, it has been shown that the critical point determined by the fidelity can be different from that obtained by using the second derivative 
of the ground-state energy \cite{jiang2018topological}.
Consequently, whether both of fidelities can describe the non-Hermitian quantum phase transitions is so far unclear.

In this paper, we clarify the puzzling problem on correct usages of the fidelity susceptibility in non-Hermitian many-body systems.
We show that the biorthogonal fidelity susceptibility instead of the self-normal fidelity susceptibility describes biorthogonal phase transitions that are associated with the gap closing.
Most importantly, we develop the perturbation theory for the fidelity susceptibility in biorthogonal bases for arbitrary interacting non-Hermitian many-body systems with real eigenvalues.
The validity of the expression is indicated with the numerical study.

This paper is organized as follows. 
In Sec.\ref{sec:PT}, we revisit the perturbation theory of the non-Hermitian systems. 
In Sec.\ref{sec:FS}, we derive the perturbative form of the biorthogonal fidelity susceptibility. 
In Sec.\ref{sec:NHTI}, we study the finite-size scaling of the non-Hermitian transverse field Ising chain. 
In Sec.\ref{sec:Con}, we summarize the results.

\section{Perturbation theory}
\label{sec:PT}
For a non-Hermitian Hamiltonian $H(\lambda)=H_0 + \lambda H^{\prime}$, where the $H(\lambda) \neq H^{\dagger}(\lambda)$, 
the eigenvalue equations of $H(\lambda)$ and $H^{\dagger}(\lambda)$ are given by \cite{brody2013biorthogonal,sternheim1972non},
\begin{align}
H(\lambda) \ket{\psi_{i}^{R}(\lambda)} = E_{i}(\lambda) \ket{\psi_{i}^{R}(\lambda)} \\
H^{\dagger}(\lambda) \ket{\psi_{i}^{L}(\lambda)} = E_{i}^{\ast}(\lambda) \ket{\psi_{i}^{L}(\lambda)}
\end{align}
Where $E_{i}(\lambda)$ are $i$th eigenvalue, and the $\ket{\psi_{i}^{L}(\lambda)}$ and $\ket{\psi_{i}^{R}(\lambda)}$ are left and right eigenvectors of the Hamiltonian $H(\lambda)$ 
that satisfies the bi-orthonormal relation \cite{brody2013biorthogonal,sternheim1972non},
\begin{align}
\braket{\psi_{i}^{L}(\lambda)} {\psi_{j}^{R}(\lambda)} = \delta_{ij}
\end{align}
and completeness relation,
\begin{align}
\sum_{i} \ket{\psi_{i}^{R}(\lambda)} \bra{\psi_{i}^{L}(\lambda)} = 1
\end{align}
In order to define a ground-state or excited states as Hermitian systems \cite{mostafazadeh2002pseudo1,mostafazadeh2002pseudo2,mostafazadeh2002pseudo3,fu2019photonic,zhao2020equivariant,chen2021quantum}, 
we assume all the eigenvalues are real, $E_{i}(\lambda) = E_{i}^{\ast}(\lambda)$, which is possible when the system has a special symmetry.
For instance, in parity-time (PT) symmetric non-Hermitian systems, the energy spectra are real in the PT symmetry unbroken regime \cite{mostafazadeh2002pseudo1,mostafazadeh2002pseudo2,mostafazadeh2002pseudo3,fu2019photonic,zhao2020equivariant,chen2021quantum}.
It is well known that the Hamiltonian  $H(\lambda)$ can be diagonalized as,
 \begin{align}
 H(\lambda) = \sum_{i} E_{i}(\lambda) \ket{\psi_{i}^{R}(\lambda)} \bra{\psi_{i}^{L}(\lambda)},
 \label{Hdiag}
 \end{align}
in biorthogonal bases.
Assuming the eigenvalues $E_{i}(\lambda)$ and the eigenvectors $\ket{\psi_{i}^{L}(\lambda)}$ and $\ket{\psi_{i}^{R}(\lambda)}$ of the Hamiltonian $H(\lambda)$ are known,
the eigenvalues $E_{i}(\lambda + \delta \lambda)$ of the Hamiltonian $H(\lambda +\delta \lambda$) can be expanded in powers of $\delta \lambda$ as \cite{sternheim1972non},
\begin{align}
E_{i}(\lambda + \delta \lambda) = E_{i}(\lambda) + \delta \lambda E_{i}^{(1)} + (\delta \lambda)^2 E_{i}^{(2)} + \cdots,
\end{align}
where $\delta \lambda \rightarrow 0$. Under the perturbation theory, the expanding coefficients $E_{i}^{(1)}$ and  $E_{i}^{(2)}$ can be derived as \cite{sternheim1972non},
  \begin{align}
  E_{i}^{(1)}  =&{} \bra{\psi_{i}^{L}(\lambda)} H^{\prime} \ket{\psi_{i}^{R}(\lambda)}, \\
  E_{i}^{(2)}  =&{} \sum_{n \neq i} \frac{ \bra{\psi_{i}^{L}(\lambda)} H^{\prime} \ket{\psi_{n}^{R}(\lambda)} \bra{\psi_{n}^{L}(\lambda)} H^{\prime} \ket{\psi_{i}^{R}(\lambda)} } {E_{i}(\lambda) - E_{n}(\lambda)}
  \label{E12}
  \end{align} 
We then have the second derivatives of ground-state energy $E_0$ per site,
  \begin{align}
  \chi_{E_{0}} =&{} \frac{1}{N} \frac{d^{2} E_{0}(\lambda) } {d\lambda^2} \label{E2def}, \\
  =&{} \frac{2}{N} E_{0}^{(2)}.
  \label{E2nd}
  \end{align} 
 Here $N$ is the system size and $d$ is the dimension of the system. We note that the $\chi_{E_{0}}$ can also be numerically obtained directly, 
 i.e. by the five-point stencil method from the ground-state energy $E_0(\lambda)$.

\begin{figure}[ht]
\includegraphics[width=8.6cm]{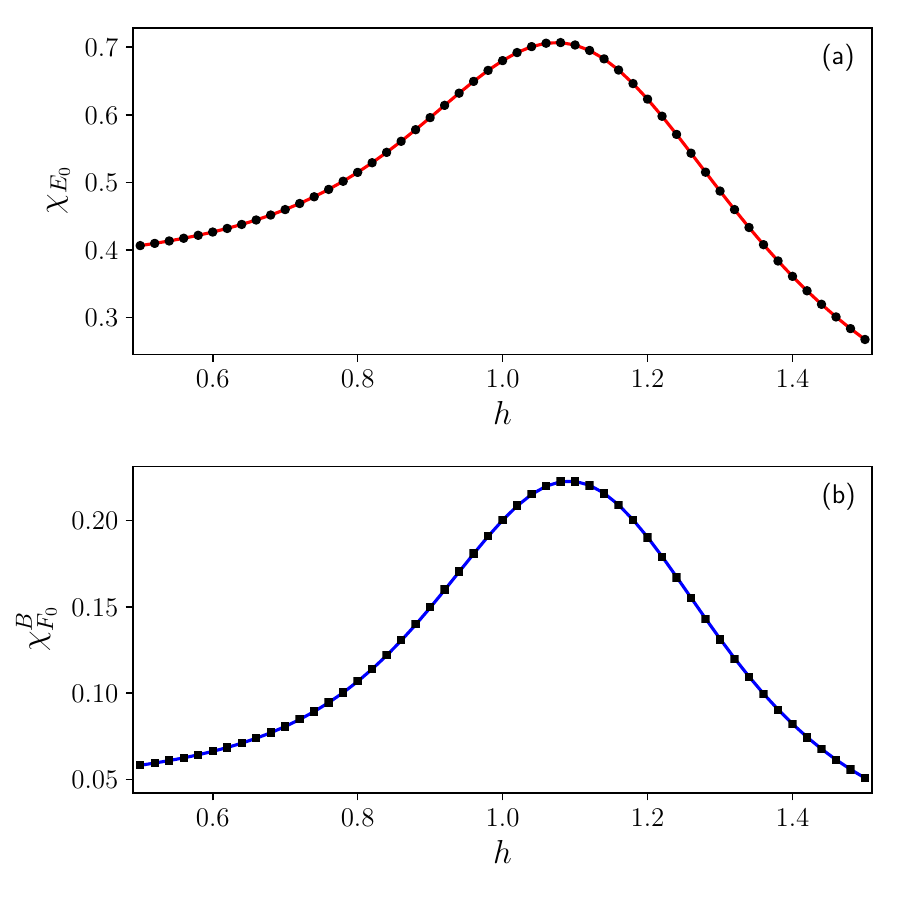} \centering
\caption{ (Color online) Perturbative results of the NHTI chain at $\gamma=0.5$ with system size $N=10$ in biorthogonal bases.
(a) Second derivatives of ground-state energy per site $\chi_{E_{0}}$, the red solid line denotes the results obtained by the five-point stencil method 
from Eq.(\ref{E2def}) with ground state energy $E_0$, the circle symbols denote the results obtained from Eq.(\ref{E2nd});
(b) Biorthogonal ground-state fidelity susceptibility per site $\chi_{F_{0}}^{B}$, the blue solid line denotes the results from Eq.(\ref{BFS}), the square symbols is given by Eq.(\ref{BFSper}).}
\label{E0FSfig}
\end{figure}

\section{Fidelity Susceptibility}
\label{sec:FS}
In this part, we develop the perturbation theory of the fidelity susceptibility. For non-Hermitian systems, we can introduce two types of fidelity susceptibility. 
First we can define a self-normal density matrix $\rho_{i}^{S}(\lambda)$ for $i$th eigenstates with only right eigenstates $\ket{\psi_{i}^{R}(\lambda)}$ 
(or only left eigenstates $\ket{\psi_{i}^{L}(\lambda)}$) as for Hermitian models, 
 \begin{align}
 \rho_{i}^{S}(\lambda) = \ket{\psi_{i}^{R}(\lambda)} \bra{\psi_{i}^{R}(\lambda)}.
 \label{SDM}
 \end{align} 
 
\begin{figure}[ht]
\includegraphics[width=8.6cm]{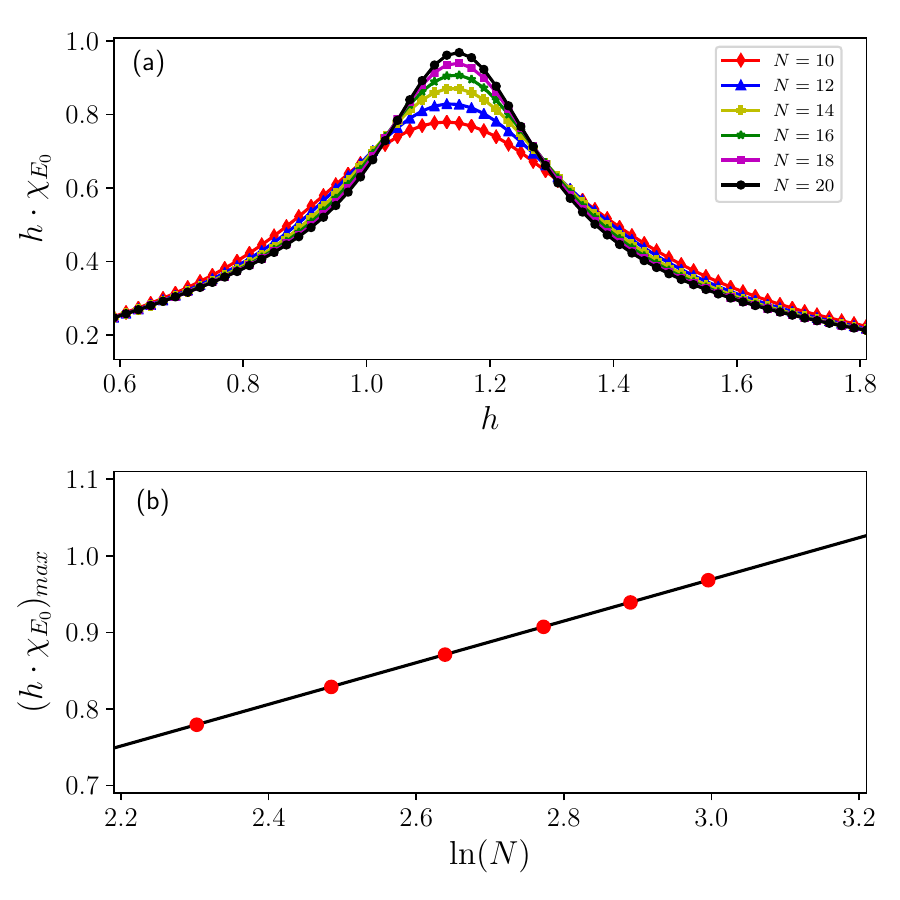} \centering
\caption{ (Color online)
Scaling of second derivatives of ground-state energy $\chi_{E_0}$ of the NHTI chain at $\gamma = 0.5$.
(a) Finite-size scaling of the $h \cdot \chi_{E_0}$ with system sizes from $N=10$ to $N=20$; 
(b) Finite-size scaling of the maxima of $h \cdot \chi_{E_0}$, where red circle symbols are the numerical results and the black solid line is the fitting curve. }
\label{E2ndfig}
\end{figure}
Here the self-normal density matrix $\rho_{i}^{S}(\lambda)$ is a Hermitian matrix, $\rho_{i}^{S \dagger}(\lambda) = \rho_{i}^{S}(\lambda)$.
However, the right eigenstates are non-orthonormal $\braket{\psi_{i}^{R}(\lambda)} {\psi_{j}^{R}(\lambda)} \neq \delta_{ij}$ due to the non-hermiticity of systems 
although each of right eigenstates can be normalized $\braket{\psi_{i}^{R}(\lambda)} {\psi_{i}^{R}(\lambda)} = 1$ independently \cite{brody2013biorthogonal}.
Alternatively, we can define a biorthogonal density matrix $\rho_{i}^{B}(\lambda)$ from Eq.(\ref{Hdiag}) by combining both right eigenstates $\ket{\psi_{i}^{R}(\lambda)}$ 
and left eigenstates $\ket{\psi_{i}^{L}(\lambda)}$ as \cite{chang2019entanglement},
 \begin{align}
 \rho_{i}^{B}(\lambda) = \ket{\psi_{i}^{R}(\lambda)} \bra{\psi_{i}^{L}(\lambda)},
 \label{BDM}
 \end{align} 
 where the biorthogonal density matrix $\rho_{i}^{B}(\lambda)$ is a non-Hermitian matrix, $\rho_{i}^{B \dagger}(\lambda) \neq \rho_{i}^{B}(\lambda)$. 
 However, left and right eigenstates satisfy the bi-orthonormal relation and the completeness relation now.
 
Consequently, the Uhlmann fidelity 
\begin{align}
F_{i} = \text{Tr} \sqrt{\sqrt{\rho_{i}(\lambda)} \rho_{i}(\lambda + \delta \lambda) \sqrt{\rho_{i}(\lambda)}}
\label{UFhi}
\end{align} 
for the self-normal density matrix $\rho_{i}(\lambda)=\rho_{i}^{S}(\lambda)$ and the biorthogonal density matrix $\rho_{i}(\lambda)=\rho_{i}^{B}(\lambda)$ can be defined 
as \cite{Gu2010,uhlmann1976transition,hauru2018uhlmann},
\begin{align}
F_{i}^{S} =&{} |\braket{\psi_{i}^{R}(\lambda)}{\psi_{i}^{R}(\lambda + \delta \lambda)}|,   \label{USF} \\
F_{i}^{B} =&{} \sqrt{\braket{\psi_{i}^{L}(\lambda + \delta \lambda)}{\psi_{i}^{R}(\lambda)} \braket{\psi_{i}^{L}(\lambda)}{\psi_{i}^{R}(\lambda + \delta \lambda)}}.  
\label{UBF}
\end{align} 
The corresponding FS per site is then given by \cite{You2007, Albuquerque2010, Gu2010,Sun2017},
\begin{align}
\chi_{F_{i}}^{S,B} = \frac{1}{N} \lim_{\delta \lambda \rightarrow 0} \frac{-2 \ln F_{i}^{S,B} }{ \delta \lambda^2}.
\label{BFS}
\end{align} 
  
We note that the perturbation theory of the self-normal fidelity susceptibility $\chi_{F_{i}}^{S}$ was recently presented in Ref.[\onlinecite{matsumoto2020continuous}]. 
A symmetric definition of the biorthogonal fidelity susceptibility $\chi_{F_{i}}^{B}$ has already been introduced in Ref.[\onlinecite{jiang2018topological}].
In this paper, we will focus mainly on the perturbation theory of biorthogonal fidelity susceptibility $\chi_{F_{i}}^{B}$ generalized from the Uhlmann fidelity.
Using the standard perturbation theory, we obtain the following perturbative form of the biorthogonal fidelity susceptibility per site in Eq.(\ref{BFS}) for $i$th eigenstates (see Appendix \ref{AppA} for details),
\begin{align}
\chi_{F_{i}}^{B} = \frac{1}{N} \sum_{n \neq i} \frac{ \bra{\psi_{i}^{L}(\lambda)} H^{\prime} \ket{\psi_{n}^{R}(\lambda)} \bra{\psi_{n}^{L}(\lambda)} H^{\prime} \ket{\psi_{i}^{R}(\lambda)} } {[E_{i}(\lambda) - E_{n}(\lambda)]^2}.
\label{BFSper}
\end{align} 
This expression is numerically checked for a non-Hermitian transversed field Ising chain as follows.

\begin{figure}[ht]
\includegraphics[width=8.6cm]{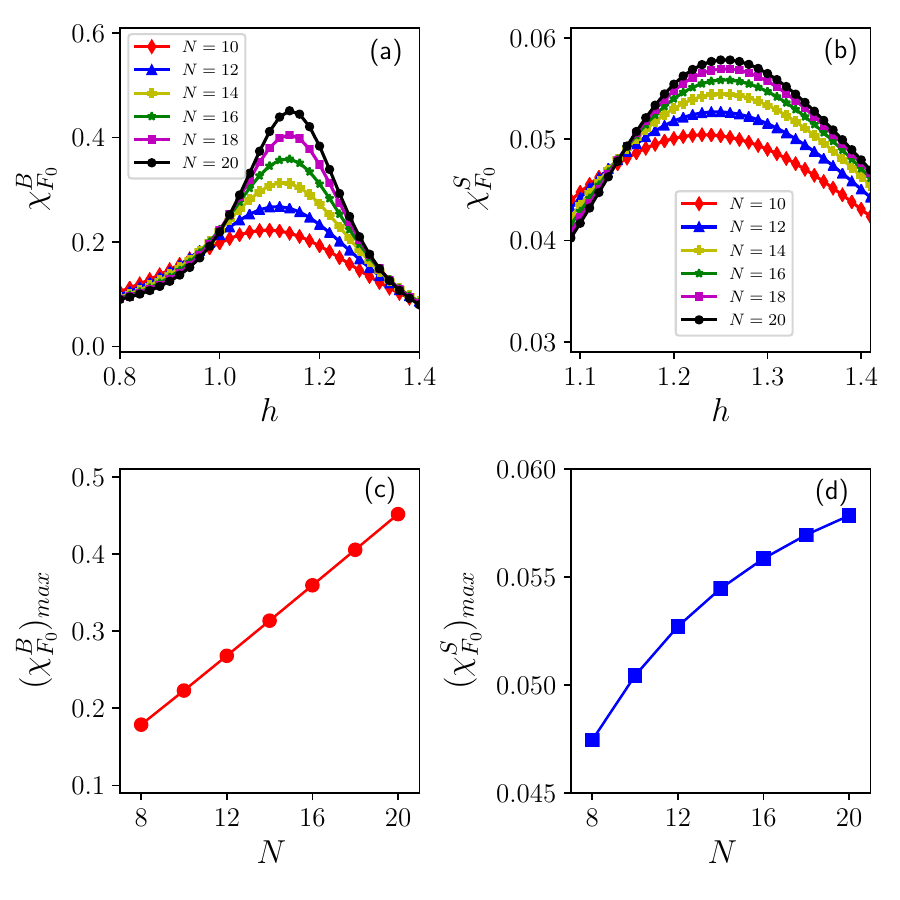} \centering
\caption{ (Color online)
Fidelity susceptibility of the NHTI chain at $\gamma = 0.5$.
(a) Biorthogonal fidelity susceptibility $\chi_{F_0}^{B}$ with respect to $h$ for system sizes from $N=10$ to $N=20$;
(b) Self-normal fidelity susceptibility $\chi_{F_0}^{S}$ as a function of $h$ with the same parameters as (a);
(c) Finite-size scaling of the maxima of $\chi_{F_0}^{B}$ in (a);
(d) Finite-size scaling of the maxima of $\chi_{F_0}^{S}$ in (b).}
\label{FSfig}
\end{figure}

\section {Model}
\label{sec:NHTI}
As an example, we consider a one-dimensional non-Hermitian transversed field Ising (NHTI) model that was studied recently in \cite{yang2020anomalous,von1991critical,bianchini2014entanglement,zhang2020topological},
 \begin{align}
 H = {}& -\sum_{j=1}^{N} J \sigma_{j}^{x} \sigma_{j+1}^{x} +  \sum_{j=1}^{N} h (\sigma_{j}^{z} + i \gamma \sigma_{j}^{y}).
 \label{HamEff}
 \end{align} 
 Here $\sigma_{j}^{x}, \sigma_{j}^{y}, \sigma_{j}^{z}$ are Pauli matrices at the $j$th site, $N$ is the number of system site.
 The coupling strength $J > 0$ and the amplitudes $h > 0$, $\gamma \geq 0$ of the transversed fields are real numbers. The $i = \sqrt{-1}$ is the imaginary unit.
 For $\gamma = 0$, the system is a Hermitian transversed field Ising model that undergoes a quantum phase transition at $h/J = 1$ between the ferromagnetic (Ferro) phase for $h/J < 1$ 
 and the paramagnetic (Para) phase for $h/J > 1$. 
 For any $\gamma \neq 0$, the system is a NHTI model because of the imaginary transverse field term along the y-axis.
The model has either all real eigenvalues for unbroken PT symmetry regimes $\gamma < 1$ 
 or complex conjugate pairs of eigenvalues for broken PT symmetry regimes $\gamma > 1$,
 with a real-complex spectral transition at $\gamma_c = 1$ (exceptional point) \cite{yang2020anomalous,zhang2020topological}. 
 We are interested in the real eigenvalues regimes ($\gamma < 1$) where the ground-state can be well defined as Hermitian models.
 In this unbroken PT symmetry regime, the system undergoes a biorthogonal order-disorder phase transition 
 between the ferromagnetic phase and the paramagnetic phase at 
 \begin{align}
 h_c = \sqrt{\frac{1}{1-\gamma^2}}
 \end{align}
 in thermodynamic limit \cite{yang2020anomalous,zhang2020topological}.
 We will focus mainly on the finite-size scaling of the ground-state fidelity susceptibility near the critical points.
 We impose periodic boundary conditions $\sigma_{N+1}^{x} =\sigma_{1}^{x}$ and use $J=1$ in our numerical simulations.
 
 We first calculate the second derivative of ground-state energy $\chi_{E_0}$ of Eq.(\ref{E2nd}) and the biorthogonal ground-state fidelity susceptibility $\chi_{F_{0}}^{B}$ of Eq.(\ref{BFSper}) 
 by performing the exact diagonalization for the NHTI model from $N=10$ to $N=20$ sizes at $\gamma=0.5$ with the step $dh = 10^{-3}$. 
 The results of $\chi_{E_0}$ and $\chi_{F_{0}}^{B}$ obtained by Eq.(\ref{E2nd}) and Eq.(\ref{BFSper}) coincide exactly with that computed from 
 the definitions in Eq.(\ref{E2def}) and Eq.(\ref{BFS}) directly [cf. Fig.\ref{E0FSfig}], indicating the perturbative formulas Eq.(\ref{E12}) and Eq.(\ref{BFSper}) we presented are valid.
 We find that the peak of second derivative of ground-state energy in the form of $h \cdot \chi_{E_0}$ increases with system sizes and diverges logarithmically [cf. Fig.\ref{E2ndfig}], 
 implying that critical exponents $\alpha = 0$ \cite{Chen2008,um2007quantum,you2009scaling}.
 
\begin{figure}[ht]
\includegraphics[width=8.6cm]{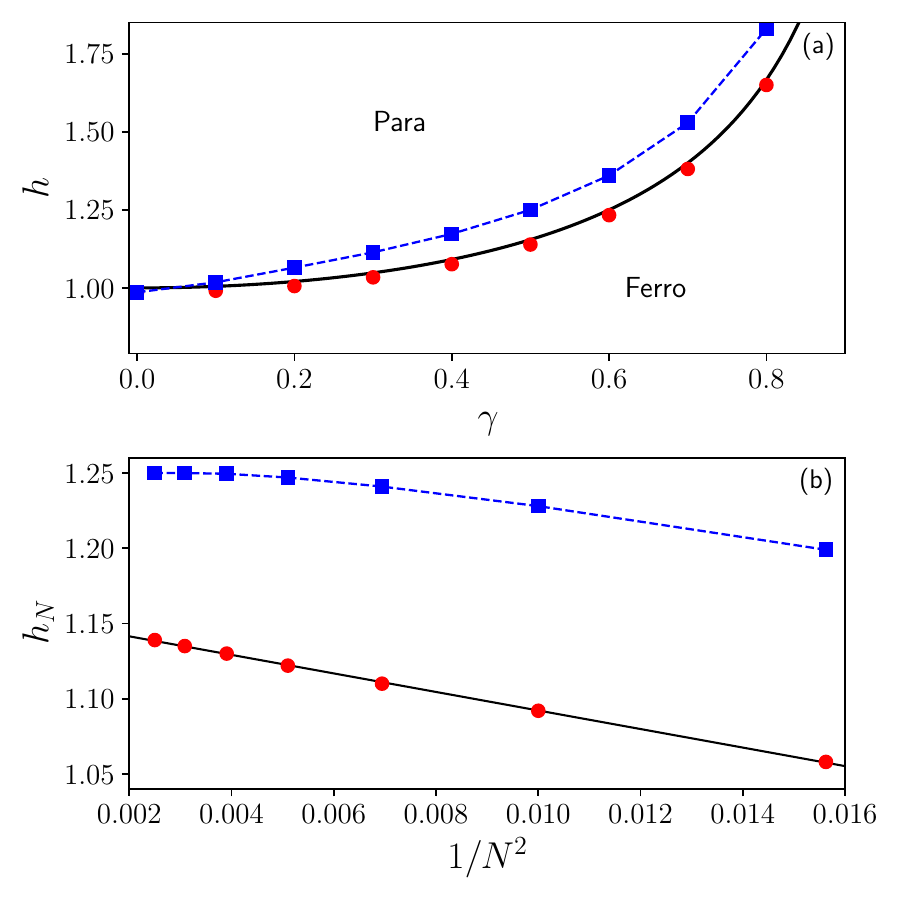} \centering
\caption{ (Color online)
Phase diagram of the NHTI chain. 
(a) Full phase diagram; Red circle symbols denote the critical values $h_N$ obtained from the biorthogonal FS $\chi_{F_0}^{B}$ for system size $N=20$; 
Blue square symbols are derived self-normal FS $\chi_{F_0}^{S}$ for system size $N=20$; the black solid line is the exact result.
(b) Blue square symbols and Red circle symbols denote the finite-size scaling of critical value $h_N$ at the maxima of the self-normal FS $\chi_{F_0}^{S}$ 
and the biorthogonal FS $\chi_{F_0}^{B}$ for $\gamma=0.5$;
the black solid line is the fitting curve with the $h_c = 1.1538$ from biorthogonal FS $\chi_{F_0}^{B}$.}
\label{phasefig}
\end{figure}

 We next discuss finite-size scaling of the biorthogonal and self-normal ground-state fidelity susceptibility $\chi_{F_{0}}^{B}$ and $\chi_{F_{0}}^{S}$ at $\gamma = 0.5$ in detail.
 As demonstrated in Fig.\ref{FSfig}, both fidelity susceptibility display a nice peak that increase with system sizes. 
 However, the finite-size scaling of $\chi_{F_{0}}^{B}$ and $\chi_{F_{0}}^{S}$ behave in a different way. 
 For biorthogonal fidelity susceptibility $\chi_{F_{0}}^{B}$, a linear scaling is found [cf. Fig.\ref{FSfig}(c)]. 
 That means we have the same correlation function critical exponents $\nu = 1$ as Hermitian transversed field Ising chain according to
 the finite-size scaling of the ground-state fidelity susceptibility \cite{You2007, Albuquerque2010, Gu2010,Sun2017}, 
 \begin{align}
 (\chi_{F_{0}}^{B})_{max} = N^{2/\nu -1}, 
 \end{align}
 for second-order phase transitions.
 For self-normal fidelity susceptibility $\chi_{F_{0}}^{S}$, a slow increase rate of the peak is observed [cf. Fig.\ref{FSfig}(d)]. 
 In addition, the critical value $h_c$ obtained from the biorthogonal FS $\chi_{F_{0}}^{B}$ tends towards 
 the exact value $h_c = 2/\sqrt{3} \approx 1.1547$ in thermodynamic limit [cf. Fig.\ref{FSfig}(a) and Fig.\ref{phasefig}(b)].
 For example, we get the critical point $h_c = 1.1538$ in thermodynamic limit for $\gamma=0.5$  [see Fig.\ref{phasefig}(b)] by extrapolating data with \cite{Damski2013}  
 \begin{align}
 h_N = h_c - a/N^2.
 \end{align}
 While the critical value $h_c$ derived from the self-normal FS $\chi_{F_{0}}^{S}$ gets worse and converges to $h_c = 1.25$ 
 when increasing the system size [cf. Fig.\ref{FSfig}(b) and Fig.\ref{phasefig}(b)].

We present the phase diagram in Fig.\ref{phasefig}(a) for $N=20$, where it is clear that the biorthogonal FS $\chi_{F_{0}}^{B}$
instead of the self-normal FS $\chi_{F_{0}}^{S}$ characterizes the biorthogonal order-disorder phase transitions. 
The critical exponents $\alpha = 0$ and $\nu=1$ derived from the finite-size scaling indicate the biorthogonal phase transitions
of the NHTI model is a second-order phase transition with the Ising universal class.

\section {Conclusion}
\label{sec:Con}
In summary, we have studied the perturbation theory of the biorthogonal fidelity susceptibility and the biorthogonal quantum criticality in interacting non-Hermitian many-body systems.
We have shown that the second derivative of ground-state energy and the biorthogonal ground-state fidelity susceptibility can serve as probes to detect quantum phase transitions 
and the corresponding critical exponents of non-Hermitian many-body systems. 
We show that the biorthogonal fidelity susceptibility instead of the conventional self-normal fidelity susceptibility should be used to characterize
phase transitions associated with the energy levels (i.e. level crossing) because the non-Hermitian Hamiltonian is diagonal in biorthogonal basis.

We note that the concept of the biorthogonal fidelity susceptibility in Eq.(\ref{BFS}) and its perturbative form as shown in Eq.(\ref{BFSper}) are general for any non-Hermitian many-body Hamiltonian with real eigenvalues. 
Consequently, it would be possible to apply the biorthogonal fidelity susceptibility to understand the nature of phase transitions in non-integrable non-Hermitian many-body models.
Moreover, it would be more interesting to know whether the biorthogonal fidelity susceptibility is useful to detect the universal class for the real-complex spectral transition 
of non-Hermitian many-body models \cite{chang2019entanglement} or the localization-delocalization transition of a non-Hermitian quantum systems \cite{hatano2021delocalization,liu2021fate,lin2021observation}
in the future.

\begin{acknowledgments}
We would like to thank M. F. Yang and W. L. You for useful discussion.
G. S. is appreciative of support from the NSFC under the Grant Nos. 11704186 and 11874220.
S. P. K  is appreciative of supported by the NSFC under the Grant Nos. 11674026, 11974053 and 12174030. 
Numerical simulations were performed on the clusters at Nanjing University of Aeronautics and Astronautics and National Supercomputing Center in Shenzhen.
\end{acknowledgments}

\appendix
\section{Perturbation theory of biorthogonal fidelity susceptibility}
\label{AppA}
Assume we know the eigenvalues $E_{i}(\lambda)$ and the left and right eigenvectors $\ket{\psi_{i}^{L}(\lambda)}$ and $\ket{\psi_{i}^{R}(\lambda)}$ of a Hamiltonian $H(\lambda)$.
According to the perturbation theory of non-Hermitian systems, the left and right eigenvectors $\ket{\psi_{i}^{L}(\lambda + \delta \lambda)}$ and $\ket{\psi_{i}^{R}(\lambda + \delta \lambda)}$ 
of the Hamiltonian $H(\lambda +\delta \lambda$) 
can be expanded in powers of $\delta \lambda$ as \cite{matsumoto2020continuous,Gu2010,Chen2008},
\begin{align}
\bra{\psi_{i}^{L}(\lambda + \delta \lambda)} = c_1 \left[ \bra{\psi_{i}^{L}(\lambda)} + \delta \lambda \sum_{n \neq i} \frac{H_{in}^{\prime} \bra{\psi_{n}^{L}(\lambda)}}{E_{i}(\lambda) - E_{n}(\lambda)} \right], \label{LW:app} \\
\ket{\psi_{i}^{R}(\lambda + \delta \lambda)} = c_2 \left[ \ket{\psi_{i}^{R}(\lambda)} + \delta \lambda \sum_{n \neq i} \frac{H_{ni}^{\prime} \ket{\psi_{n}^{R}(\lambda)}}{E_{i}(\lambda) - E_{n}(\lambda)} \right],
\label{RW:app}
\end{align} 
up to the first order. Where $H_{ni}^{\prime}=\bra{\psi_{n}^{L}(\lambda)} H^{\prime} \ket{\psi_{i}^{R}(\lambda)}$, 
$c_1=\braket{\psi_{i}^{L}(\lambda + \delta \lambda)}{\psi_{i}^{R}(\lambda)}$ and $c_2=\braket{\psi_{i}^{L}(\lambda)}{\psi_{i}^{R}(\lambda + \delta \lambda)}$ are the normalization constants.
We can get the biorthogonal fidelity susceptibility $F_{i}^{B}$ in terms of the $c_1$ and $c_2$ by multiplying equation (\ref{LW:app}) by right eigenvectors $\ket{\psi_{i}^{R}(\lambda)}$ 
and multiplying equation (\ref{RW:app}) by the left eigenvectors $\bra{\psi_{i}^{L}(\lambda)}$ respectively,
 \begin{align}
 (F_{i}^{B})^2 =&{} \braket{\psi_{i}^{L}(\lambda + \delta \lambda)}{\psi_{i}^{R}(\lambda)} \braket{\psi_{i}^{L}(\lambda)}{\psi_{i}^{R}(\lambda + \delta \lambda)} \nonumber \\
 =&{} c_1c_2
 \label{BFS:app}
 \end{align} 
 Multiplying equation (\ref{LW:app}) by equation (\ref{RW:app}) and using the normalization condition
 $ \braket{\psi_{i}^{L}(\lambda + \delta \lambda)}{\psi_{i}^{R}(\lambda +\delta \lambda)} = 1,$
we derive the equation of biorthogonal fidelity,
\begin{align}
1 = (F_{i}^{B})^2 \left[ 1 + (\delta \lambda)^2 \sum_{n \neq i} \frac{H_{in}^{\prime} H_{ni}^{\prime}}{[E_{i}(\lambda) - E_{n}(\lambda)]^2} \right].
\end{align} 
Where the Eq.(\ref{BFS:app}) has been used. The biorthogonal fidelity susceptibility per site can be obtained as,
\begin{align}
\chi_{F_{i}}^{B} = \frac{1}{N} \sum_{n \neq i} \frac{ \bra{\psi_{i}^{L}(\lambda)} H^{\prime} \ket{\psi_{n}^{R}(\lambda)} \bra{\psi_{n}^{L}(\lambda)} H^{\prime} \ket{\psi_{i}^{R}(\lambda)} } {[E_{i}(\lambda) - E_{n}(\lambda)]^2}.
\end{align} 
by considering the leading term to second-order. 

\section{Differential form of biorthogonal fidelity susceptibility}
\label{AppB}
Next we will derive the differential form of the biorthogonal FS $\chi_{F_{i}}^{B}$ for the $i$th state. 
The left and right eigenvectors $\ket{\psi_{i}^{L}(\lambda + \delta \lambda)}$ and $\ket{\psi_{i}^{R}(\lambda + \delta \lambda)}$ of the Hamiltonian $H(\lambda +\delta \lambda$) 
are firstly expanded using Taylor series in powers of $\delta \lambda$ as \cite{matsumoto2020continuous,Gu2010,Chen2008},
\begin{align}
\bra{\psi_{i}^{L}(\lambda + \delta \lambda)} ={}& \bra{\psi_{i}^{L}(\lambda)} + \delta \lambda \bra{\partial_\lambda \psi_{i}^{L}(\lambda)} \nonumber \\
 {}&~~ + \frac{\delta \lambda^2}{2} \bra{\partial^{2}_{\lambda} \psi_{i}^{L}(\lambda)} +O(\delta \lambda^3), \label{LW2:app} \\
\ket{\psi_{i}^{R}(\lambda + \delta \lambda)} ={}& \ket{\psi_{i}^{R}(\lambda)} +\delta \lambda \ket{\partial_\lambda \psi_{i}^{R}(\lambda)} \nonumber \\
 {}&~~ + \frac{\delta \lambda^2}{2} \ket{\partial^{2}_{\lambda} \psi_{i}^{R}(\lambda)} +O(\delta \lambda^3),
\label{LRW:app}
\end{align} 
Hence the overlap $\braket{\psi_{i}^{L}(\lambda + \delta \lambda)}{\psi_{i}^{R}(\lambda)}$ and $\braket{\psi_{i}^{L}(\lambda)}{\psi_{i}^{R}(\lambda + \delta \lambda)}$ are given as,
\begin{align}
\braket{\psi_{i}^{L}(\lambda + \delta \lambda)}{\psi_{i}^{R}(\lambda)} ={}& 1 + \delta \lambda \braket{\partial_\lambda \psi_{i}^{L}(\lambda)}{\psi_{i}^{R}(\lambda)} \nonumber \\
{}&~~ + \frac{\delta \lambda^2}{2} \braket{\partial_\lambda^2 \psi_{i}^{L}(\lambda)}{\psi_{i}^{R}(\lambda)} \\
\braket{\psi_{i}^{L}(\lambda)}{\psi_{i}^{R}(\lambda + \delta \lambda)} ={}& 1 + \delta \lambda \braket{\psi_{i}^{L}(\lambda)}{\partial_\lambda \psi_{i}^{R}(\lambda)} \nonumber \\
{}&~~ + \frac{\delta \lambda^2}{2} \braket{\psi_{i}^{L}(\lambda)}{\partial_\lambda^2 \psi_{i}^{R}(\lambda)},
\label{LRW2:app}
\end{align} 
Where the bi-orthonormal relation $\braket{\psi_{i}^{L}(\lambda)}{\psi_{i}^{R}(\lambda)}=1$ is used. From Eq.(\ref{BFS:app}), we have 
\begin{align}
(F_{i}^{B})^2 ={}& \braket{\psi_{i}^{L} (\lambda + \delta \lambda)}{\psi_{i}^{R}(\lambda)} \braket{\psi_{i}^{L}(\lambda)}{\psi_{i}^{R}(\lambda + \delta \lambda)} \nonumber \\
   ={}& 1 +  \delta \lambda \left[ \braket{\partial_\lambda \psi_{i}^{L}(\lambda)}{\psi_{i}^{R}(\lambda)} + \braket{\psi_{i}^{L}(\lambda)}{\partial_\lambda \psi_{i}^{R}(\lambda)} \right] \nonumber \\
      {}& ~~ +  \frac{\delta \lambda^2} {2} \left[ 2 \braket{\partial_\lambda \psi_{i}^{L}(\lambda)}{\psi_{i}^{R}(\lambda)} \braket{\psi_{i}^{L}(\lambda)}{\partial_\lambda \psi_{i}^{R}(\lambda)} \right. \nonumber \\
      {}& ~~~~~~~~ + \left. \braket{\partial_\lambda^2 \psi_{i}^{L}(\lambda)}{\psi_{i}^{R}(\lambda)} + \braket{\psi_{i}^{L}(\lambda)}{\partial_\lambda^2 \psi_{i}^{R}(\lambda)} \right] 
\label{UBF2}
\end{align} 
up to the second order of $\delta \lambda^2$. From the bi-orthonormal relation $\braket{\psi_{i}^{L}(\lambda)}{\psi_{i}^{R}(\lambda)}=1$, we can get
\begin{align}
\partial_{\lambda} \braket{\psi_{i}^{L}(\lambda)}{\psi_{i}^{R}(\lambda)} ={}& \braket{\partial_\lambda \psi_{i}^{L}(\lambda)}{\psi_{i}^{R}(\lambda)} + \braket{\psi_{i}^{L}(\lambda)}{\partial_\lambda \psi_{i}^{R}(\lambda)} \nonumber \\
={}& 0 \label{RL1} \\
\partial_{\lambda}^2 \braket{\psi_{i}^{L}(\lambda)}{\psi_{i}^{R}(\lambda)} ={}& \braket{\partial_{\lambda}^2 \psi_{i}^{L}(\lambda)}{\psi_{i}^{R}(\lambda)} + \braket{\psi_{i}^{L}(\lambda)}{\partial_{\lambda}^2 \psi_{i}^{R}(\lambda)} \nonumber \\
{}& ~~ + 2 \braket{\partial_\lambda \psi_{i}^{L}(\lambda)}{\partial_\lambda \psi_{i}^{R}(\lambda)} \nonumber \\
={}& 0
\label{RL2}
\end{align} 
Using the relations Eq.(\ref{RL1}) and Eq.(\ref{RL2}), the Eq.(\ref{UBF2}) becomes
\begin{align}
(F_{i}^{B})^2 ={}& 1 - \delta \lambda^2 N \chi_{F_{i}}^{B},
\end{align}
where the biorthogonal FS per site $\chi_{F_{i}}^{B}$ is defined as
\begin{align}
\chi_{F_{i}}^{B} = \frac{1}{N} {}& \left[ \braket{\partial_\lambda \psi_{i}^{L}(\lambda)}{\partial_\lambda \psi_{i}^{R}(\lambda)}  \right. \nonumber \\
{}& - \left. \braket{\partial_\lambda \psi_{i}^{L}(\lambda)}{\psi_{i}^{R}(\lambda)} \braket{\psi_{i}^{L}(\lambda)}{\partial_\lambda \psi_{i}^{R}(\lambda)} \right].
\end{align}

\bibliographystyle{apsrev4-1}
\bibliography{ref}

\end{document}